\documentclass[11pt]{article}
\usepackage{xspace}
\usepackage{graphicx}
\usepackage{amsmath}
\usepackage{amssymb}
\usepackage{color}

\definecolor{Red}{rgb}{1,0,0}
\definecolor{Green}{rgb}{0,1,0}
\definecolor{Blue}{rgb}{0,0,1}
\definecolor{Black}{rgb}{0,0,0}

\def\beq{\begin{equation}}
\def\eeq#1{\label{#1}\end{equation}}
\def\eeqn{\end{equation}}

\def\beqa{\begin{eqnarray}}
\def\eeqa#1{\label{#1}\end{eqnarray}}
\def\eeqan{\end{eqnarray}}

\let\bar=\overbar

\def\Dslash{\not{\hbox{\kern-4pt $D$}}}
\def\dslash{\not{\hbox{\kern-2pt $\del$}}}

\def\msb{{\bar{\ssstyle M \kern -1pt S}}}

\def\Title#1{\begin{center} {\Large {\bf #1} } \end{center}}

  \let\OLDthebibliography\thebibliography
\renewcommand\thebibliography[1]{
  \OLDthebibliography{#1}
  \setlength{\parskip}{0pt}
  \setlength{\itemsep}{0pt plus 0.3ex}
}

\begin{document}

\Title{A combined limit for neutrinoless double-beta decay}

\bigskip\bigskip

\begin{raggedright}  

{\it Pawel Guzowski\index{Guzowski, P.},\\
School of Physics and Astronomy\\
University of Manchester\\
Manchester M13 9PL, UK}\\

\end{raggedright}
\vspace{1.cm}

{\small
\begin{flushleft}
\emph{To appear in the proceedings of the Prospects in Neutrino Physics Conference, 15 -- 17 December, 2014, held at Queen Mary University of London, UK.}
\end{flushleft}
}

\section{Introduction}

The search for neutrinoless double-beta decay is important
in determining the Majorana nature of the neutrino, and also
in establishing if lepton number is violated. The half-life $T_{1/2}^{0\nu}$
for this process is given by
\begin{equation}
  [T_{1/2}^{0\nu}]^{-1} = G^{0\nu} |M^{0\nu}|^2 \frac{m_{\beta\beta}^2}{m_e^2} \; , \label{eqn:halflife}
\end{equation}
where $G^{0\nu}$ is a phase space 
factor and $M^{0\nu}$ a nuclear matrix element (NME) factor (both these factors depend on the nuclear isotope), $m_e$ the electron mass, and $m_{\beta\beta}$ the effective neutrino mass.

In recent years five experiments have published results of their searches: CUORICINO (observing $^{130}$Te) \cite{cuore}, EXO ($^{136}$Xe) \cite{exo}, GERDA ($^{76}$Ge) \cite{gerda}, KamLAND-Zen ($^{136}$Xe) \cite{kamland}, and NEMO-3 ($^{100}$Mo) \cite{nemo3}. There has been no previous attempt to systematically combine the limits of their searches using their measured energy distributions. Presented here is the first such combination, divided into two sections: first a cross check of the individual experimental results, showing that the method can replicate their limits, and then the combination of all five experiments.

The results of this analysis are published in Ref.~\cite{combipaper}.

\section{Individual experiments}

\begin{figure}[!ht]
\begin{center}
\includegraphics[width=0.45\columnwidth]{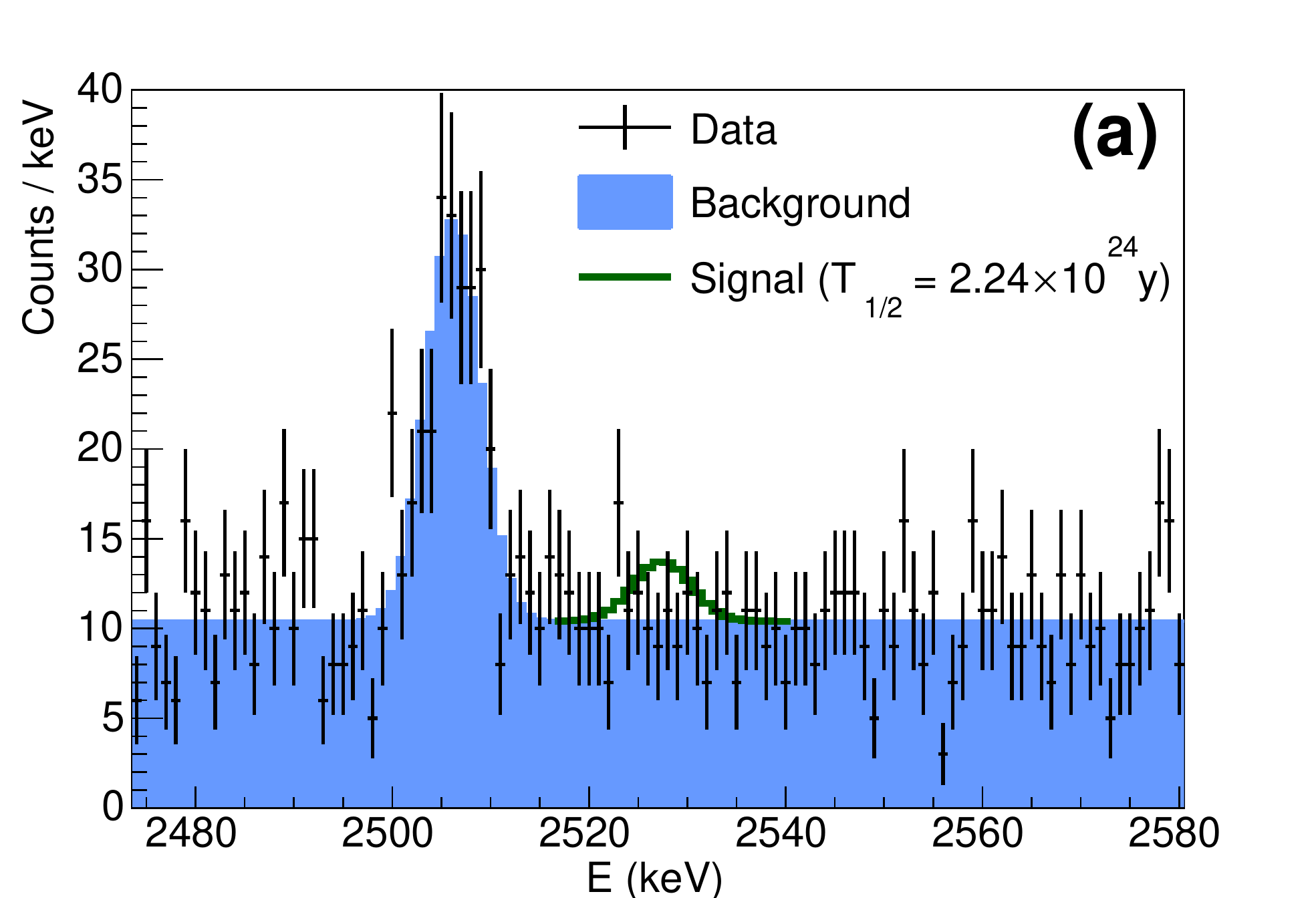}
\includegraphics[width=0.45\columnwidth]{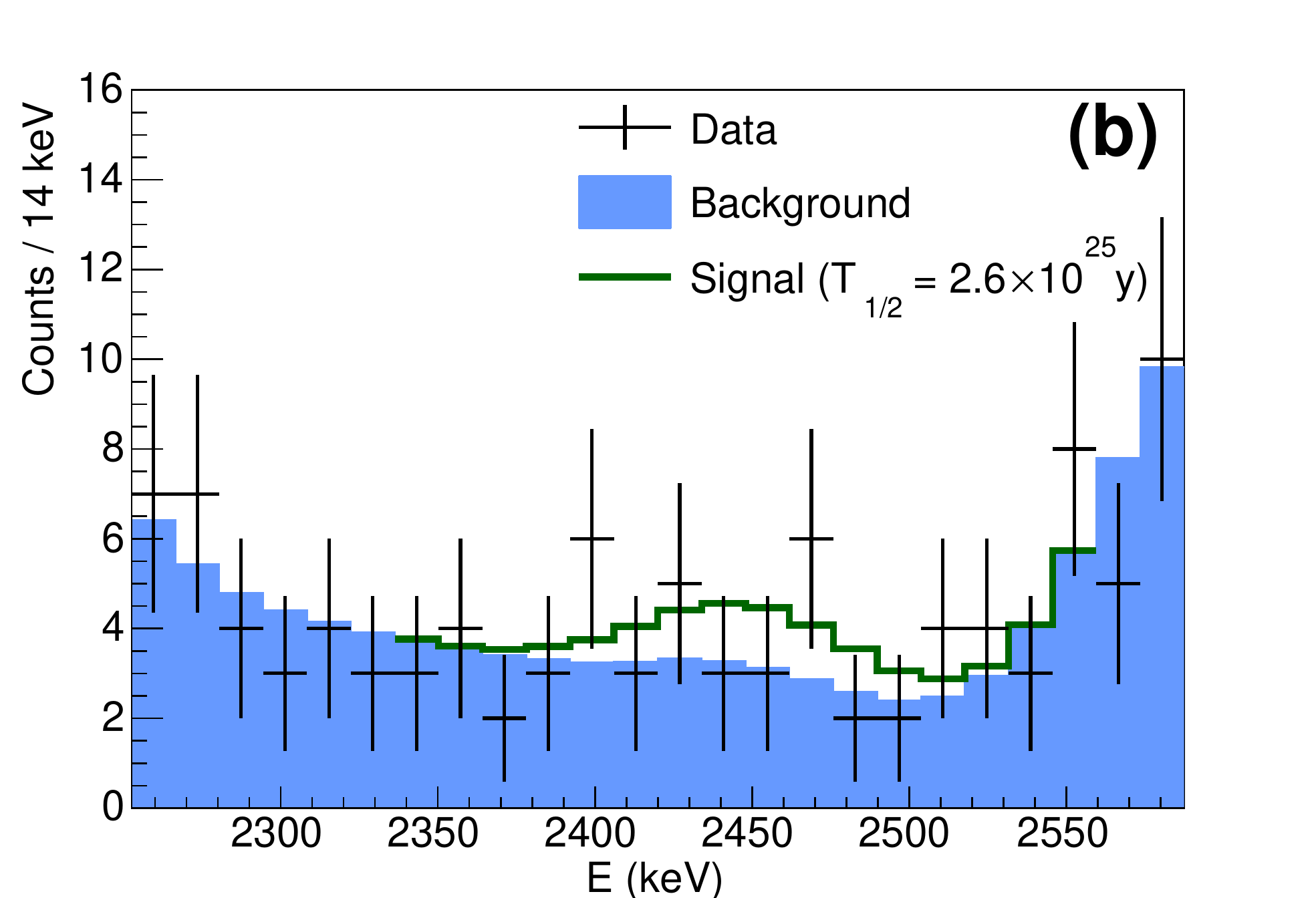}
\includegraphics[width=0.45\columnwidth]{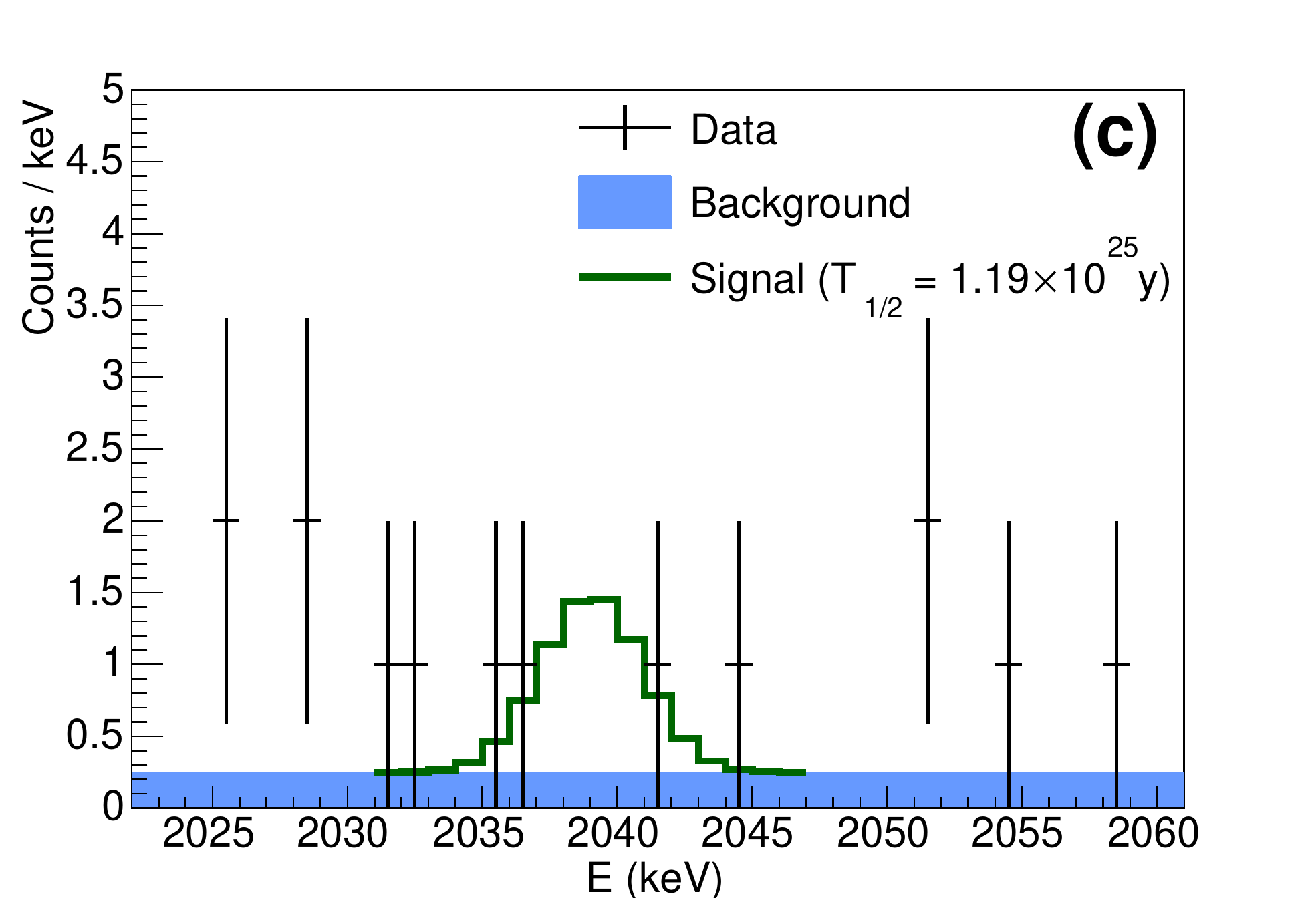}
\includegraphics[width=0.45\columnwidth]{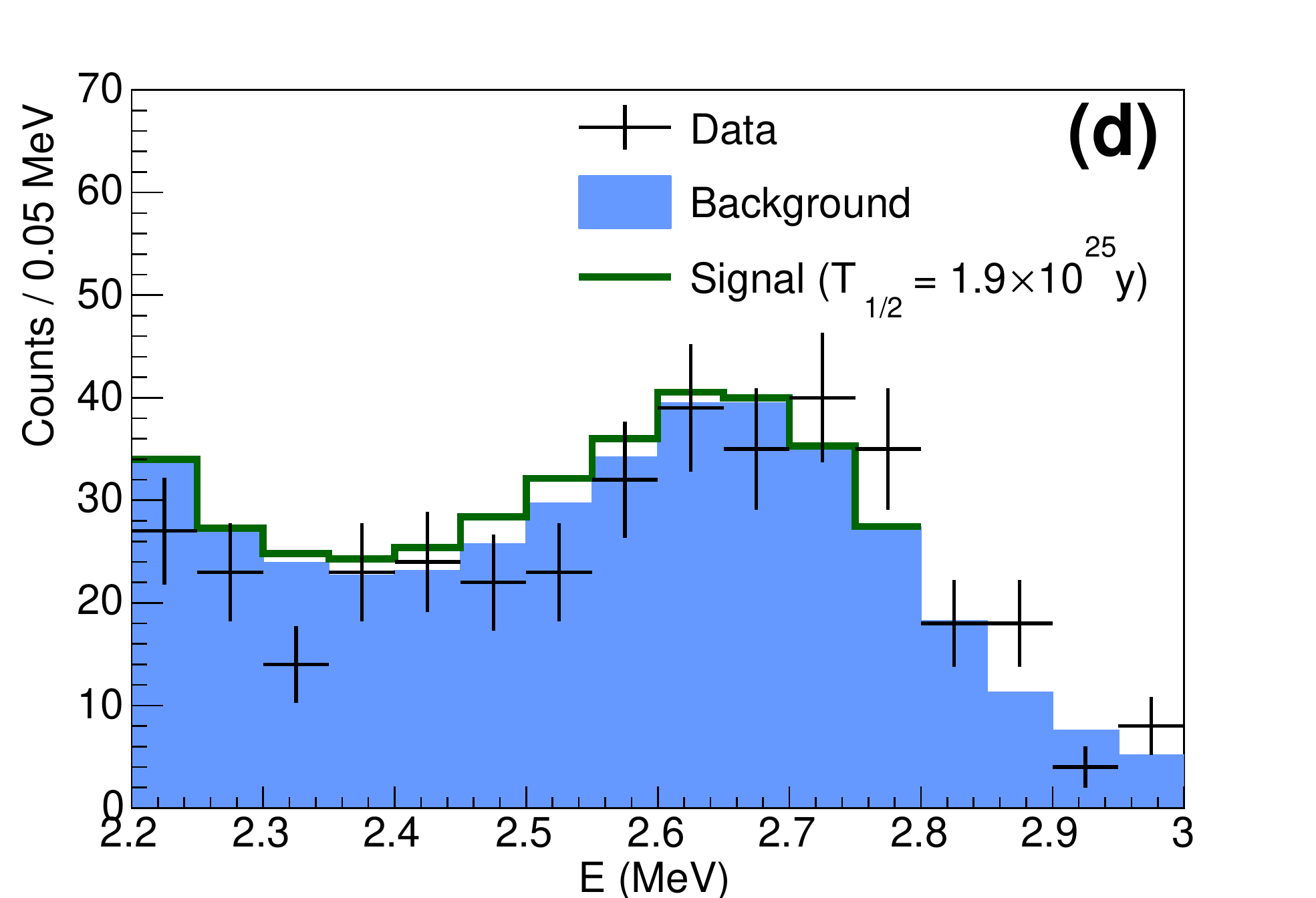}
\includegraphics[width=0.45\columnwidth]{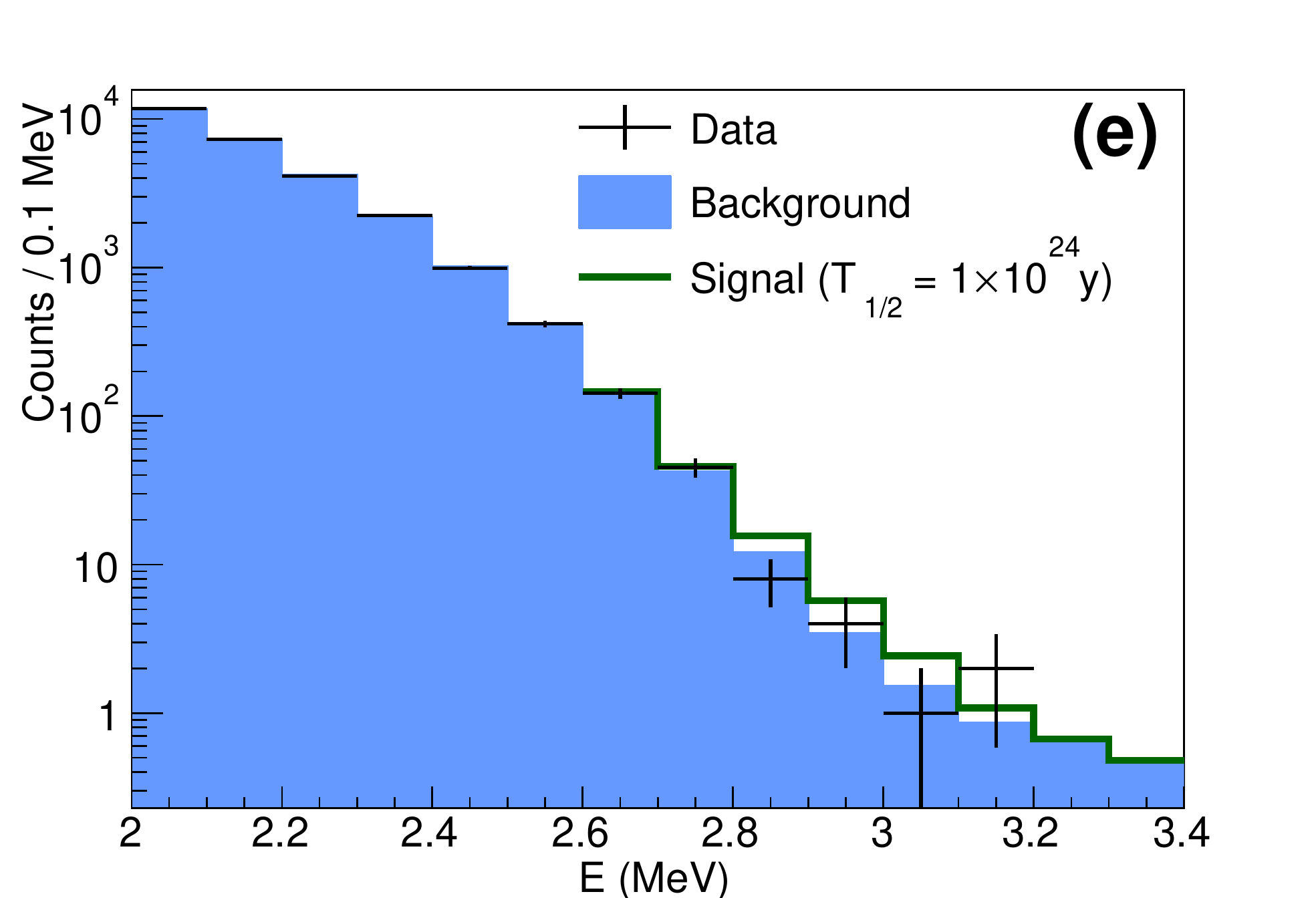}
\caption{The input distributions for (a) CUORICINO (fig. 9 of \cite{cuore}); (b) EXO (fig. 4(a) of \cite{exo}); (c) GERDA (fig. 1 of \cite{gerda}); (d) KamLAND-Zen (fig. 1(a) of \cite{kamland}); and (e) NEMO-3 (fig. 2 of \cite{nemo3}).}
\label{fig:inputs}
\end{center}
\end{figure}
The input background, signal and data energy distributions for each experiment, along with the systematic uncertainties, are taken from their publications \cite{cuore,exo,gerda,kamland,nemo3}, and are shown in Figure~\ref{fig:inputs}.
\begin{table}[!ht]
  \begin{center}
    \caption{The published limits 
on $T^{0\nu}_{1/2}$ for each experiment, and the calculated observed and 
expected (median $\pm$ $1\sigma$) limits.}
\begin{tabular}{lcccc }
\hline\hline
 & \multicolumn{4}{c}{90\% CL lower limit on $T^{0\nu}_{1/2}~(10^{24}$ y)}  \\
\cline{2-5}
 Experiment    & Published & Observed & Expected & $\pm 1 \sigma$ range  \\
\hline
 CUORICINO    & 2.8  & 2.8      & 2.9      & 2.0 -- 4.2        \\
 EXO        & 11     & 13       & 21       & 14  -- 30        \\
 GERDA      & 21    & 20       & 21       & 14  -- 29         \\
 KamLAND-Zen  & 19    & 17       & 11       & \phantom{0}7   -- 15        \\
 NEMO-3      & 1.1    & 1.1      & 0.9      & 0.6 -- 1.4         \\
\hline\hline
\end{tabular}
\label{tab:replicate}
\end{center}
\end{table}
Using these distributions and the signal normalisations, the observed 90\% CL lower half-life limits can be calculated, and are shown in Table~\ref{tab:replicate}, along with the experiments' published limits, and the expected limits from these distributions (both median expected and the $\pm1\sigma$ ranges). Overall the calculated limits agree well with the published limits, validating both the method and the use of these distributions in the combination.

\section{Combination}

As the experiments use different isotopes, their half-life limits cannot be directly combined. Instead the limits are calculated for a common $m_{\beta\beta}$ and translated to the signal normalisation in each experiment's energy distribution using Equation~\ref{eqn:halflife}, with the values of $G^{0\nu}$ and $M^{0\nu}$ taken from recent calculations \cite{phasespace,GCM,IBM-2,NSM,QRPA,RQRPA}. The limit is calculated for each NME model seperately.
\begin{table}[!ht]
  \begin{center}
    \caption{The combined observed and expected $m_{\beta\beta}$ upper limits, the improvements of the combination with respect to the best individual experiment ((E)XO, (G)ERDA or (K)amLAND-Zen) for that NME model, and the $p$ value for the combined limit with respect to the Heidelberg-Moscow (HM) positive claim~\cite{kk}.}
\begin{tabular}{lccccc} 
      \hline\hline
        NME model    & $m^{\rm obs}_{\beta\beta}$  & $m^{\rm exp}_{\beta\beta}$  &    \multicolumn{2}{c}{Improvement} & $p$ value\\
           &  (meV) & (meV) &  Limit  & Sensitivity &   (HM)\\
      \hline
      GCM \cite{GCM} & 130 & 130 & 12\% (K) & 8\% (E) & 0.001\\
      IBM-2 \cite{IBM-2} & 190 &  180 & 15\% (K) & 12\% (E) & 0.023\\
      NSM \cite{NSM} & 310 & 290 & 13\% (K) & 10\% (E) & 0.004\\
      QRPA \cite{QRPA} & 190 & 180 & 26\% (K)  & 22\% (E)  & 0.095\\
      (R)QRPA \cite{RQRPA} & 300 & 300 & 20\% (G) & 19\% (G) & 0.39\\
      \hline\hline
   \end{tabular}
\label{tab:limits}
\end{center}
\end{table}
Table~\ref{tab:limits} shows the 90\% CL upper limit on $m_{\beta\beta}$ fo each NME calculation, and also shows the improvement in the mass limit over the best individual experiment in that NME model. There can be an improvement of up to $\sim25\%$ in the mass limit, which corresponds to an increase of exposure of a factor of up to $\sim3$. The combined mass limit ranges from $130-310$~meV, depending on the NME model.
\begin{figure}[!ht]
\begin{center}
\includegraphics[width=0.6\columnwidth]{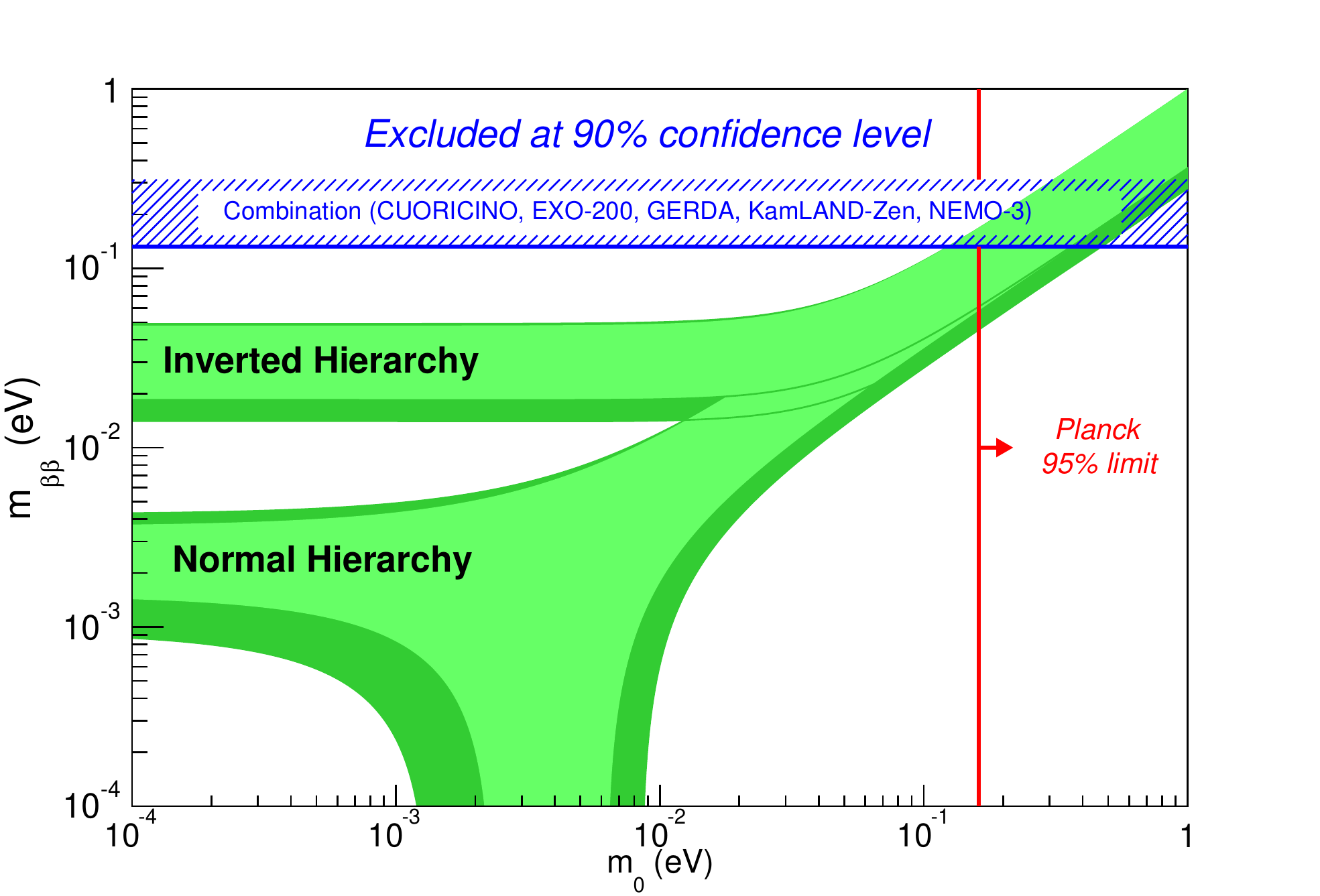}
\caption{The combined $m_{\beta\beta}$ limit range overlaid on the range of allowed $m_{\beta\beta}$ for a given mass $m_0$ of the lightest neutrino mass eigenstates in the normal and inverted mass hierarchies. Also shown is the range of $m_0$ disfavoured by cosmology.}
\label{fig:mBB_m0}
\end{center}
\end{figure}
This limit is shown in Figure~\ref{fig:mBB_m0}, in the context of the allowed $m_{\beta\beta}$ regions depending on the neutrino mass hierarchy and the lightest neutrino mass eigenstate.

We also compare these mass limits (translated back into half-life limits of $^{76}$Ge) to the positive claim for neutrinoless double-beta decay of $^{76}$Ge in the Heidelberg-Moscow experiment, which measuread $T^{0\nu}_{1/2}=(2.23^{+0.44}_{-0.31}) \times 10^{25}$~y~\cite{kk}. The level of compatability of the limits with respect to this claim varies strongly with the NME model used, with $p$~values ranging from 0.001--0.4.

\section{Summary}

A first direct combination of neutrinoless double-beta decay experiments on multiple isotopes has been performed, yielding an upper limit on the effective neutrino mass of $130-310$~meV. The combination can improve the limits on the effective neutrino mass by up to $25\%$ compared to the best individual experiment, depending on the choice of NME model. The compatibility of the combined limits with respect to the claimed observation of neutrinoless double-beta decay in the Heidelberg-Moscow 
experiment also varies significantly depending on the NME calculations chosen.


\begin{thebibliography}{99}
\bibitem{cuore}
  E. Andreotti {\it et al.} (CUORICINO Collaboration), Astropart.\ Phys.\ {\bf 34} (2011) 822.
\bibitem{exo}
  J. B. Albert {\it et al.} (EXO-200 Collaboration), Nature {\bf 510} (2014) 229.
\bibitem{gerda}
  M.~Agostini {\it et al.} (GERDA Collaboration), Phys.\ Rev.\ Lett.\ {\bf 111} (2013) 122503.
\bibitem{kamland}
  A. Gando {\it et al.} (KamLAND-Zen~Collaboration), Phys.\ Rev.\ Lett.\ {\bf 110} (2013) 062502.
\bibitem{nemo3}
R. Arnold {\it et al.} (NEMO-3 Collaboration),  Phys.\ Rev.\ D {\bf 89} (2014) 111101(R)
\bibitem{combipaper}
  P.~Guzowski, L.~Barnes, J.~Evans, G.~Karagiorgi, N.~McCabe, and S.~S\"oldner-Rembold, arXiv:1504.03600 [hep-ex] (2015), Submitted to Phys. Rev. D.
\bibitem{phasespace}
J.~Kotila and F.~Iachello, Phys. Rev. C {\bf 85} (2012) 034316.
\bibitem{GCM}
  T.~R. Rodriguez and G.~Martinez-Pinedo, Phys.\ Rev.\ Lett.\ {\bf 105} (2010) 252503.
\bibitem{IBM-2}
  J. Barea, J. Kotila, and F. Iachello, Phys. Rev. C {\bf 91} (2015) 034304.
 \bibitem{NSM}
J.~Menendez, A.~Poves, E.~Caurier, and F.~Nowacki, Nucl. Phys. A {\bf 818} (2009) 139.
\bibitem{QRPA}
F.~{\v S}imkovic, V.~Rodin, A.~Faessler, and P.~Vogel, Phys. Rev. C {\bf 87} (2013) 045501.
\bibitem{RQRPA}
A.~Faessler, G.~Fogli, E.~Lisi, V.~Rodin, A.~Rotunno, and F.~{\v{S}}imkovic, Phys. Rev. D {\bf 79} (2009) 053001.
\bibitem{kk}
H.~Klapdor-Kleingrothaus and I.~Krivosheina,  Mod.\ Phys.\ Lett.\ A {\bf 21} (2006) 1547.
\end{thebibliography}
\end{document}